\documentstyle[preprint,aps]{revtex}
\begin{document}
\title{Physics Potential of the e-RHIC Based FEL-Nucleus Collider }
\author{H. Koru$^{a}$, A. \"{O}zcan$^{a}$, S. Sultansoy$^{a,b}$and \"{O}. Yava\c{s}$%
^{c}$}
\address{$^{a}$Physics Dept., Faculty of Science and Arts,\\
Gazi University, 06500\\
Teknikokullar, Ankara, TURKEY\\
$^{b}$Institute of Physics, Academy of Sciences, H. Cavid Ave. 33, Baku,\\
AZERBAIJAN\\
$^{c}$Dept. of Physics Engineering, Faculty of Engineering, Ankara\\
University, 06100 Tando\u{g}an, Ankara, TURKEY}
\maketitle

\begin{abstract}
Main parameters of the FEL-RHIC collider are estimated. The physics search
potential of this machine is illustrated using excitations of the $^{232}$Th
nucleus as an example. It is shown that, due to the tunability and
monochromacity of FEL beam and high statistics, proposed collider will play
an important role in the field of ''traditional'' nuclear physics.
\end{abstract}

\bigskip

Recently, construction of a 10 GeV energy electron linac tangentially to
Relativistic Heavy Ion Collider (RHIC, BNL, USA) is proposed\cite{I.Ben-Zvi}
in order to investigate the physics of collisions between electrons and
heavy ions, and between polarized electrons and polarized protons \cite{Work}%
. Let us cite one paragraph from introduction part of the Ref.\cite
{I.Ben-Zvi}:

''There is interest among the synchrotron light community in the
construction of a superconducting linac in the energy range from 6 GeV to 10
GeV, in order to drive a free electron laser. Except when RHIC is in the
(relatively rare) process of cooling down from ambient temperatures, the
cryogenic refrigation plant already installed has enough spare capacity to
run such a linac. It is therefore natural to place the linac relatively near
to RHIC. The synergy is completed by making spare pulses from the linac
available for eRHIC collisions.''

In this letter we are arguing that collisions of the Free Electron Laser
(FEL) photons with ultra-relativistic nuclei from RHIC will give a unique
possibility to investigate nucleus excitations. In estimations below we
assume that photon beam is produced by SASE FEL based on 10 GeV TESLA-like
linac, taking in mind successful results achieved at the TESLA Test Facility
(TTF) Free-Electron Laser at the DESY\cite{J.An}.

A general schematic view of FEL-RHIC collider is given in Fig 1. The photon
beam produced by undulator magnet placed at the end of 10 GeV e-linac
collides with the nucleus beam from RHIC. Relativistic ions will \ ''see''
FEL beam as a \ ''laser'' photons with energy $\omega =2\gamma _{A}\omega
_{0}$, where $\gamma _{A}$ is the Lorentz factor of ions and $\omega _{0}$\
is the energy of the FEL photons. The FEL-RHIC is a good candidate to become
a pioneer of FEL-nucleus colliders, to be followed by FEL-HERA and FEL-LHC 
\cite{H.Aktaþ}. The tunability of $\omega _{0}$ by changing the electron
beam energy will give opportunity for resonant production of nucleus
excitations. The excited nucleus will turn to the ground state at a distance 
$\ell =\gamma _{A}\cdot \tau \cdot c$ from collision point, where $\tau $ is
the lifetime of the excited state in the nucleus rest frame and c is the
speed of light. Emitted photons will be seen in detector as high energy
photons with energies up to $2\gamma _{A}E_{\gamma },$ where $E_{\gamma }$
is the energy of emitted photons in the rest frame of the nucleus.

The wavelength of the first harmonic of SASE FEL radiation is given by

\begin{equation}
\lambda =\frac{\lambda _{u}}{2\gamma _{e}^{2}}(1+\frac{k^{2}}{2})
\end{equation}
where $\lambda _{u}$ is the period length of an undulator, $\gamma
_{e}=E_{e}/m_{e}$ is the Lorentz factor of the electron, $k=eB_{u}\lambda
_{u}/2\pi m_{e}$ is the undulator parameter, $e=\sqrt{4\pi \alpha }$ ($%
\alpha $ being the fine structure constant) and $B_{u}$ is the peak value of
the magnetic field in the undulator. At the TTF FEL with $E_{e}=235$ $MeV$
the observed SASE FEL photons has the wavelength 109 nm, which corresponds
to  $\omega _{0}=11$ $eV.$ According to Eq. (1), 10 GeV linac based FEL will
produce photons with the wavelength $0.06$ $nm,$ which corresponds to $%
\omega _{0}=20$ $keV.$ On the other hand, the wavelength ot the FEL photons
is constrained by $\lambda >4\pi \varepsilon _{t}$, where $\varepsilon
_{t}=\varepsilon _{N}/\gamma _{e}$ and $\varepsilon _{N\text{ \ }}$is
normalized emittance of the electron beam. For e-linac based FEL's, $%
\varepsilon _{N}$ value is usually around 10$^{-6}$ $m\cdot rad$, hence $%
\lambda >0.6$ $nm$ for $E_{e}=10$ $GeV$. \ In order to achieve shorter
wavelenghts one should either use ''cooled'' electron beam with smaller
emittance or deal with higher FEL harmonics. The last case leads to decrease
of the luminosity. For example, the brighteness of the third harmonic (which
has 3 times shorter wavelength) is approximately 1\% of the first harmonic
one \cite{LCLS}. Combining both methods, the FEL\ photon energy of 20 keV
and even more can be achieved at FEL-RHIC Collider.

Parameters of $^{232}Th$ beam are presented in Table I, where we use the
parameters of gold nucleus for linac-ring option of e-RHIC as reference
point \cite{I.Ben-Zvi}. Parameters of SASE FEL beam produced by a 10 GeV
TESLA-like linac \cite{R.Brink} are given in Table II. Luminosity of the
FEL-nucleus collisions is given by

\begin{equation}
L=\frac{n_{\gamma }n_{A}}{4\pi \sigma _{x}^{eff}\sigma _{y}^{eff}}%
n_{b}f_{rep}
\end{equation}
where $n_{\gamma }$ and $n_{A}$ are the number of photons and nuclei in the
FEL and nucleus bunches, respectively, $n_{b}$ is the number of electron
bunches per train, $f_{rep}$ is TESLA repetition rate, $\sigma _{x}^{_{eff}}$
and $\sigma _{y}^{_{eff}}$ are the values which are obtained by choosing the
bigger ones of the corresponding horizontal and vertical sizes of the FEL
and the nucleus beams (for round beams $\sigma _{x}=\sigma _{y}$). Using
parameters given in Tables I and II we obtain $L=$ $2\cdot 10^{30}$ $%
cm^{-2}s^{-1}$ with $n_{\gamma }=10^{13}$.

Today, more than hundred excitations of the $^{232}$Th nucleus are founded
in different experiments \cite{Nuclear} and only seven of them are observed
in ($\gamma $ ,$\gamma ^{^{\prime }}$) reactions. The reasons are following 
\cite{U.Kneis}: low statistics, limited energy resolution, neutron induced
background from electron setup material collisions etc. In Table III we
present main characteristics of $^{232}$Th excitations with low spin $(J\leq
2)$.

The cross section for the resonant photon scattering is given by the
well-known Breit-Wigner formula

\begin{equation}
\sigma \left( \gamma ,\gamma ^{\prime }\right) =\frac{\pi }{E^{2}}\cdot 
\frac{2J_{exc}+1}{2\left( 2J_{0}+1\right) }\cdot \frac{B_{in}B_{out}\Gamma
^{2}}{\left( E-E_{R}\right) ^{2}+\Gamma ^{2}/4}
\end{equation}
where $E$ is the c.m. energy of the incoming photon (in our case it is very
close to that in the rest frame of the nucleus), $J_{exc}$ and $J_{0}$ are
spins of the excited and ground states of the nucleus, $B_{in}$ and $B_{out}$
are the branching fractions of the excited nucleus into the entrance and
exit channels, respectively, $E_{R}$ is the energy at the resonance and $%
\Gamma $ is the total width of the excited nucleus. Corresponding resonant
cross sections are given in Table III.

The energy of FEL photons needed for excitation of corresponding $^{232}$Th
level can be expressed as:

\begin{equation}
\omega _{0}=\frac{E_{exc}}{2\gamma _{Th}}
\end{equation}
where $E_{exc}$ is the energy of corresponding excited level, $\gamma _{Th}$
is the Lorentz factor of the $^{232}$Th nucleus. Corresponding values for
needed FEL photon energies are given in Table IV. Taking into account the
energy spread of the FEL and the nucleus beam ($\Delta E_{\gamma }/E_{\gamma
}=\allowbreak \Delta \omega /\omega _{0}=10^{-4}$ and $\Delta
E_{A}/E_{A}=10^{-4}$), the approximate value of averaged cross section has
been found to be

\begin{equation}
\sigma _{av}\approx \sigma _{res}\frac{\Gamma }{\Delta E_{\gamma }}
\end{equation}
where $\Delta E_{\gamma }\approx 10^{-4}E_{exc}$. Corresponding values for
averaged cross sections are presented in the third column of Table IV.
Multiplying these values with the luminosity of FEL-RHIC colliders ($2\cdot
10^{30}$ $cm^{-2}s^{-1})$, we obtain number of events per second which are
given in the last column of Table IV. As seen typical numbers of events are
of the order of millions per minute, whereas ''traditional'' experiments
deals with hundreds events.

FEL-RHIC colliders will give opportunity to ''measure'' unknown decay width
using known ones. Indeed, decay width can be estimated using following
relation

\begin{equation}
\Gamma_{\left( 1\right) }\cong \frac{E_{\gamma }^{\left( 1\right) }} {%
E_{\gamma }^{\left( 2\right) }}\cdot \frac{N_{\left( 1\right) }} {%
N_{\left(2\right) }}\cdot \frac{\sigma_{res}^{\left( 2\right) }} {%
\sigma_{res}^{\left( 1\right) }}\cdot \Gamma_{\left( 2\right) }  \label{1}
\end{equation}
where index 1 (2) corresponds to level with unknown (known) decay width. Let
us use the 49.368 keV level as known one and 100 events per second as
observation limit. In this case we will be able to ''measure'' decay width
of the 714.25 (1078.7) keV level at FEL -Th RHIC collider, if it exceeds 2.5 
$\cdot $10$^{-7}$ (2.6$\cdot $10$^{-6}$) eV.

We had mentioned that the excited nucleus will turn to the ground state at a
distance $\ell =\gamma _{A}\cdot \tau \cdot c$ from collision point and the
emitted photons will be seen in detector as high energy photons with
energies up to $2\gamma _{A}E_{\gamma }$ . As an example, $\ell $ is equal
to 0.46 (8.2$\cdot $10$^{-4}$) $m$ for the 49.369 (2296) keV excitation of $%
^{232}$Th nucleus at RHIC. Therefore, the detector could be placed close to
the collision region. The photons with 49.369 (2296) keV emitted in the rest
frame of the nucleus will be seen in the detector as high-energy photons
with energies up to 10.1 (470) MeV.

In fixed target experiments the spin of excited nucleus can be determined
using angular distribution of the emitted photons. In our case, this angular
distribution will be transferred to the energy distribution in laboratory
frame. For spin 1 and 2 cases, angular distributions in the rest frame are
given by \cite{U.Kneis} 
\begin{equation}
W(\theta )=\frac{3}{4}(1+\cos ^{2}\theta )\bigskip \   \label{2}
\end{equation}
\begin{equation}
W(\theta )=\frac{5}{4}(1-3\cos ^{2}\theta +4\cos ^{4}\theta )  \label{3}
\end{equation}
respectively. In laboratory system, these distributions will be seen by
detector as energy distributions (for $\gamma >>1$) 
\begin{equation}
W(x)=\frac{3}{4}(x^{2}-2x+2)  \label{4}
\end{equation}
\begin{equation}
W(x)=\frac{5}{4}(4x^{4}-16x^{3}+21x^{2}-10x+2)  \label{5}
\end{equation}
where $x=E_{\gamma }/\gamma _{A}\omega .$ Here, $x$ varies from 0 to 2 ($x=0$
corresponds to $\theta =180^{0}$ and $x=2$ corresponds to $\theta =0^{0}$ ).
Fig. 2 shows the $x$ dependence of normalized energy distributions. Taking
into account the high statistics, provided by proposed experiment, it is
obvious that different spin values can be easily identified.

In the nucleus rest frame, parity $\pi $ of spin 1 dipole excitations in a
nucleus with 0$^{+}$ground state can be determined \cite{N.Piet} using
linearly polarized FEL beam by measuring 
\begin{eqnarray}
\Sigma  &=&\frac{W(\theta =90^{0},\varphi =0^{0})-W(\theta =90^{0},\varphi
=90^{0})}{W(\theta =90^{0},\varphi =0^{0})+W(\theta =90^{0},\varphi =90^{0})}
\nonumber \\
&=&\pi _{1}=\left\{ 
\begin{array}{clc}
+1, & \mbox{for} & J^{\pi }=1^{+}, \\ 
-1, & \mbox{for} & J^{\pi }=1^{-}.
\end{array}
\right. 
\end{eqnarray}
In our case, $\theta =90^{0}$ corresponds to photons with $E_{\gamma
}=\gamma _{A}\omega ,$ which are emitted at $\eta _{\max }=1/\gamma _{A},$
where $\eta $ is the angle between emitted photon and initial nucleus beam
direction. Azimuthal angle $\varphi $ with respect to the (horizontal)
polarization plane of the $\gamma $ beam is unchanged when transferred to
the laboratory frame. If the detector is placed at a distance 100 $m$ from
the interaction point, the spot size of emitted photons will have a radius
about 1 $m$. Therefore, the measurement of parity can be made easily.

Finally, let us compare proposed experiment with existing ones. As mentioned
in Ref. \cite{U.Kneis}:

''There are several methods to produce photons for low energy photon
scattering experiments ... An ideal photon source for such experiments
should have the following characteristics:

$\cdot $ High spectral intensity $I=N_{\gamma }/eV\cdot s$ (number of
photons per energy bin and second),

$\cdot $ Good monochromaticity $\Delta E_{\gamma }/E_{\gamma }$

$\cdot $ Tunable in a broad energy range,

$\cdot $ High degree of linear polarization ($P_{\gamma }\approx $ 100\%).

Up to now there are no such ideal sources available fulfilling all these
requirements in every respect. Therefore, diverse photon sources have been
applied in low energy photon scattering depending on the special
experimental requirements and aims intended in the investigations ...''

FEL-RHIC collider fulfills all these requirements! This statement is
illustrated by Table V, where characteristics of the different photon
sources are presented. Moreover, since the accelerated nuclei are fully
ionized, the background which is induced by low-shell electrons in the case
of investigation of excitations of heavy nuclei using traditional methods
will be eliminated.

In conclusion, we hope that the huge number of events provided by FEL-RHIC
collider and pure experimental environment will give opportunity to
investigate the most of known excitations of $^{232}$Th nucleus ($\sim $100
levels), which are observed by different experiments, in $\left( \gamma
,\gamma ^{\prime }\right) $ reactions, too, as well as to observe a lot of
additional levels. Similar analysis can be easily performed for other
nuclei. In general FEL-RHIC collider will essentially enlarge the role of $%
(\gamma ,\gamma ^{\prime })$reactions in nuclear physics research.

We are grateful to A.K. \c{C}ift\c{c}i and Y.\.{I}slamzade for useful
discussions. This work is partially supported by Turkish State Planning
Organization under the Grant No 2002 K 120250.

\bigskip 
%TCIMACRO{
%\TeXButton{B}{\begin{table}[tbp] \centering%
%}}%
%BeginExpansion
\begin{table}[tbp] \centering%
%
%EndExpansion
\caption{Parameters of $^{232}$Th beam\label{key}}

\begin{tabular}{|l|r|}
\hline
Lorentz factor & 104 \\ \hline
Normalized emittance $\left( \pi \mu m\right) $ & 6 \\ \hline
Bunch population $\left( 10^{9}\right) $ & 1.7 \\ \hline
Ring circumference $\left( m\right) $ & 3833 \\ \hline
Bunches per ring $\left( k\right) $ & 180 \\ \hline
Bunch spacing $\left( ns\right) $ & 71 \\ \hline
RMS beam size at the IP $\left( \mu m\right) $ & 60 \\ \hline
\end{tabular}

\bigskip 
%TCIMACRO{
%\TeXButton{E}{\end{table}%
%}}%
%BeginExpansion
\end{table}%
%
%EndExpansion

%TCIMACRO{
%\TeXButton{B}{\begin{table}[tbp] \centering%
%}}%
%BeginExpansion
\begin{table}[tbp] \centering%
%
%EndExpansion
\caption{Parameters of a 10 GeV TESLA -like FEL beam\label{key}}

\begin{tabular}{|l|r|}
\hline
Linac repetition rate $\left( Hz\right) $ & 5 \\ \hline
Number of bunches per train $\left( k\right) $ & 11000 \\ \hline
Photon energy range $\left( keV\right) $ & 0.1$\div $ 20 \\ \hline
Number of photons per bunch $\left( 10^{12}\right) $ & 1$\div $ 100 \\ \hline
Photon beam divergence $\left( \mu rad\right) $ & 1 \\ \hline
Photon beam diameter $\left( \mu m\right) $ & 20 \\ \hline
\end{tabular}

\bigskip\ 
%TCIMACRO{
%\TeXButton{E}{\end{table}%
%}}%
%BeginExpansion
\end{table}%
%
%EndExpansion

%TCIMACRO{
%\TeXButton{B}{\begin{table}[tbp] \centering%
%}}%
%BeginExpansion
\begin{table}[tbp] \centering%
%
%EndExpansion
\caption{Main characteristics of some of the $^{232}$Th nucleus
excitations\label{key}}

\begin{tabular}{|c|c|c|c|c|c|}
\hline
$E_{exc}$ $\left( keV\right) $ & $J^{\pi }$ & $\left( \gamma ,\gamma
^{\prime }\right) $ & $T_{1/2}$ $\left( s\right) $ & $\Gamma $ $\left(
eV\right) $ & $\sigma _{res}$ $\left( cm^{2}\right) $ \\ \hline
49.369 & $2^{+}$ & + & $345\cdot 10^{-12}$ & $1.91\cdot 10^{-6}$ & $%
5.02\cdot 10^{-18}$ \\ \hline
714.25 & $1^{-}$ & + & $-$ & $-$ & $1.43\cdot 10^{-20}$ \\ \hline
774.1 & $2^{+}$ & - & $5.8\cdot 10^{-12}$ & $1.13\cdot 10^{-4}$ & $2.03\cdot
10^{-20}$ \\ \hline
785,3 & $2^{+}$ & - & $2.7\cdot 10^{-12}$ & $2.43\cdot 10^{-4}$ & $1.97\cdot
10^{-20}$ \\ \hline
1053.6 & $2^{+}$ & - & $-$ & - & $1.09\cdot 10^{-20}$ \\ \hline
1072.9 & $2^{+}$ & - & $-$ & - & $1.06\cdot 10^{-20}$ \\ \hline
1077.5 & $1^{-}$ & - & $-$ & - & $6.30\cdot 10^{-21}$ \\ \hline
1078.7 & $0^{+}$ & + & $-$ & $-$ & $2.10\cdot 10^{-21}$ \\ \hline
1121.8 & $2^{+}$ & - & $-$ & - & $9.69\cdot 10^{-21}$ \\ \hline
1387.2 & $2^{+}$ & - & $1.4\cdot 10^{-12}$ & $4.70\cdot 10^{-4}$ & $%
6.34\cdot 10^{-21}$ \\ \hline
1489.3 & $(1,2^{+})$ & - & $-$ & - & $(3.30,5.50)\cdot 10^{-21}$ \\ \hline
1554.2 & $2^{+}$ & - & $2.95\cdot 10^{-12}$ & $2.23\cdot 10^{-4}$ & $%
5.05\cdot 10^{-21}$ \\ \hline
1561.5 & $(1,2^{+})$ & - & $-$ & - & $(3.00,5.00)\cdot 10^{-21}$ \\ \hline
1573.0 & $(1,2^{+})$ & - & $-$ & - & $(2.96,4.93)\cdot 10^{-21}$ \\ \hline
1578.5 & $2^{+}$ & - & $-$ & - & $4.90\cdot 10^{-21}$ \\ \hline
1738.1 & $(1,2^{+})$ & - & $-$ & - & $(2.92,4.04)\cdot 10^{-21}$ \\ \hline
2043 & $1^{+}$ & + & $8.64\cdot 10^{-15}$ & $0.076$ & $1.75\cdot 10^{-21}$
\\ \hline
2248 & $1^{+}$ & + & $1.8\cdot 10^{-14}$ & $0.037$ & $1.45\cdot 10^{-21}$ \\ 
\hline
2274 & $1^{+}$ & + & $3.9\cdot 10^{-14}$ & $0.017$ & $1.41\cdot 10^{-21}$ \\ 
\hline
2296 & $1^{+}$ & + & $2.64\cdot 10^{-14}$ & $0.025$ & $1.39\cdot 10^{-21}$
\\ \hline
\end{tabular}

\bigskip 
%TCIMACRO{
%\TeXButton{E}{\end{table}%
%}}%
%BeginExpansion
\end{table}%
%
%EndExpansion

\bigskip 
%TCIMACRO{
%\TeXButton{B}{\begin{table}[tbp] \centering%
%}}%
%BeginExpansion
\begin{table}[tbp] \centering%
%
%EndExpansion
\caption{The $^{232}$Th excitations at the FEL--RHIC
collider\label{key}}\ 
\begin{tabular}{|c|c|c|c|}
\hline
$E_{exc}$ $\left( keV\right) $ & $\omega _{0}$ $\left( eV\right) $ & $\sigma
_{av}$ $\left( cm^{2}\right) $ & $N_{events}/s$ \\ \hline
$49.369$ & $237$ & $1.94\cdot 10^{-24}$ & $3.88\cdot 10^{6}$ \\ \hline
$714.25$ & $3433$ & $-$ & $-$ \\ \hline
$774.1$ & $3721$ & $2.96\cdot 10^{-26}$ & $5.92\cdot 10^{4}$ \\ \hline
$785.3$ & $3775$ & $6.09\cdot 10^{-26}$ & $1.22\cdot 10^{5}$ \\ \hline
1053.6 & $5065$ & - & - \\ \hline
1072.9 & 5158 & - & - \\ \hline
1077.5 & 5180 & - & - \\ \hline
$1078.7$ & $5186$ & $-$ & $-$ \\ \hline
1121.8 & 5393 & - & - \\ \hline
$1387.2$ & $6669$ & $2.15\cdot 10^{-26}$ & $4.30\cdot 10^{4}$ \\ \hline
1489.3 & 7160 & - & - \\ \hline
$1554.2$ & $7472$ & $7.24\cdot 10^{-27}$ & $1.45\cdot 10^{4}$ \\ \hline
1561.5 & 7507 & - & - \\ \hline
1573.0 & 7562 & - & - \\ \hline
1578.5 & 7588 & - & - \\ \hline
1738.1 & 8356 & - & - \\ \hline
$2043$ & $9822$ & $6.51\cdot 10^{-25}$ & $1.30\cdot 10^{6}$ \\ \hline
$2248$ & $10807$ & $2.39\cdot 10^{-25}$ & $4.78\cdot 10^{5}$ \\ \hline
$2274$ & $10932$ & $1.05\cdot 10^{-25}$ & $2.10\cdot 10^{5}$ \\ \hline
$2296$ & $11038$ & $1.51\cdot 10^{-25}$ & $3.02\cdot 10^{5}$ \\ \hline
\end{tabular}

\bigskip 
%TCIMACRO{
%\TeXButton{E}{\end{table}%
%}}%
%BeginExpansion
\end{table}%
%
%EndExpansion

%TCIMACRO{
%\TeXButton{B}{\begin{table}[tbp] \centering%
%}}%
%BeginExpansion
\begin{table}[tbp] \centering%
%
%EndExpansion
\caption{Characteristics of the different photon sources (for details and
corresponding references see [8]).\label{key}}

\begin{tabular}{|l|l|l|l|l|l|}
\hline
Photon Source & $LCP$ & $BS_{p}$ & $BS_{up}+CP$ & $BS_{up}$ & $FEL-RHIC$ \\ 
\hline
Spectral Intensity($\gamma $/s$\cdot $eV) & $0.15$ & $20$ & $1000$ & $1000$
& $10^{16}\cdot MeV/E_{exc}$ \\ \hline
$\Delta $E$_{\gamma }$ /E$_{\gamma }$ (\%) & $2.7$ & $cont.$ & $cont.$ & $%
cont.$ & $0.01$ \\ \hline
P$_{\gamma }$ (\%) & $100$ & $10-30$ & $10-20$ & $0$ & $100$ \\ \hline
Target Mass M (g) & $70$ & $5$ & $5$ & $1-2$ & $10^{-10}$ \\ \hline
\end{tabular}

\bigskip 
%TCIMACRO{
%\TeXButton{E}{\end{table}%
%}}%
%BeginExpansion
\end{table}%
%
%EndExpansion

\end{document}